\documentclass[%
 reprint,
 amsmath,amssymb,
 aps,
]{revtex4-2}

\usepackage{graphicx}
\usepackage{dcolumn}
\usepackage{bm}
\usepackage[nice]{nicefrac}

\begin{document}

\title{Fluctuating interfaces in barotropic beta-plane turbulence}
\author{Sandip Sahoo}
 \email{sandipsahoo09902@gmail.com}
\author{Samriddhi Sankar Ray}%
\email{samriddhisankarray@gmail.com}
\affiliation{%
 International Centre for Theoretical Sciences, Tata Institute of Fundamental Research, Bangalore 560089, India
}%

\date{\today}

\begin{abstract}

	Zonal jets manifest themselves as bands with sharp interfaces in the
	vorticity configuration. We develop an algorithm to track these
	fluctuating vorticity interfaces and systematically investigate their
	characteristic spatio-temporal behavior. While the interfacial height
	fluctuations are typically sub-Gaussian, the corresponding
	\textit{fluctuation speeds} exhibit wider, heavy-tailed distributions
	reflecting the influence of lateral dispersion induced by the zonal
	velocity profile along the interfacial contours. The temporal evolution
	of these fluctuations is further characterized through their power
	spectrum displaying scale invariance in the frequency domain. The
	sharp, dense, shock-like features present in the time series of the
	\textit{height} field suggest a possible lacking of differentiability.
	We confirm this by calculating the moments of the time-increments of
	the interfacial height fluctuations.  Finally, the fractal nature of
	these boundaries is investigated systematically through a multifractal
	approach, revealing the non-trivial, complex statistics of interfaces
	in such geophysical, turbulent flows. 

\end{abstract}

\maketitle

Fully developed turbulent flows are commonly seen in nature and
industry. Quite often, studies in turbulence rely on assumptions of
homogeneity and isotropy, which is still a valid approximation for a
large number of phenomena. Importantly, the framework of fully
developed, statistically homogeneous and isotropic
turbulence---starting with the seminal work of Kolmogorov in
1941---remains the cornerstone in our understanding of the
mathematical and statistical physics structure of high Reynolds
number flows~\cite{Frisch_1995}. This is especially true as many of the tools of
statistical physics, used to address such fundamental questions as
the universality of intermittency and anomalous scaling, rest on
such notions of (statistical) isotropy and homogeneity~\cite{MukherjeePRL2024}.

\begin{figure*}
	\centering
	\includegraphics[width=1.0\linewidth]{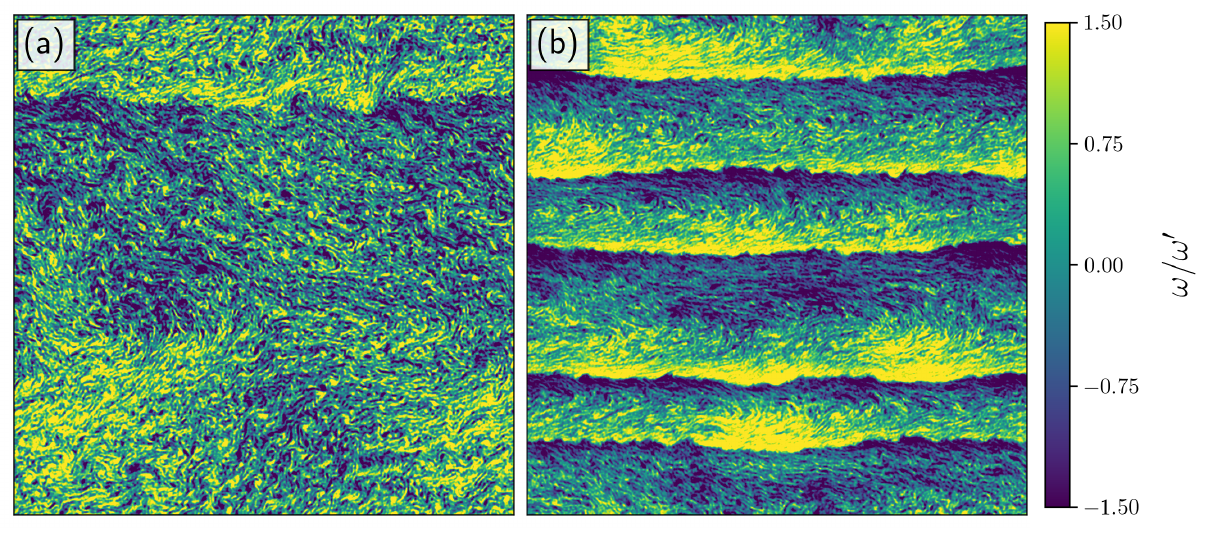}
	\caption{Pseudocolor plots of the vorticity fields with the zonostrophy parameter 
		(a) $R_\beta = 6.3$ ($\beta = 10$) and (b) $R_\beta = 7.5$ ($\beta = 50$). Distinct jets, a key signature of zonal turbulence, 
		are clearly seen which become more prominent for the case of larger $R_\beta$. A time evolution 
		of these flows, seen in an animation~\protect \cite{YT1}, reveal the rich spatio-temporal dynamics of the vorticity interfaces.}
	\label{fig:illus}
\end{figure*}

Nevertheless, there are several instances of turbulent flows in nature
where the lack of statistical isotropy or homogeneity is apparent.
The most striking natural examples of such turbulence are zonal flows,
commonly observed in several planetary atmospheres as well as in
terrestrial oceans~\cite{Kondra, dePater_Lissauer_2015, galperin2019zonal}. They are also a common feature of strongly magnetised plasmas in tokamaks and solar tachocline~\cite{Stringer_1969, Rozhansky1996, Diamond_2005}. Zonal flows---with characteristic, distinct,
banded, jet-like structures (see Fig.~\ref{fig:illus}(b) for a
representative plot)---in geophysical systems emerge from an interplay of the Coriolis
force and the inherent turbulence driven by the high Reynolds numbers
typical of these systems~\cite{Rhines_1975, Galperin}.  Perhaps the
most striking manifestations of such flows and their characteristic
banded structure are observed in the Jovian atmosphere as well as  in
most of the outer planets such as Saturn, Uranus and Neptune~\cite{Bolton_2017, galperin2019zonal, Read_2024, Dowling_2019}.	

On Earth, such zonal turbulent flows have immediate consequences. The
Coriolis force acting on atmospheric circulation gives rise to
alternating jets of easterlies and westerlies moving anti-parallel to
each other~\cite{Hadley_1735, Lewis_1998, Uppala_2005}, which become apparent only after time-averaging. The
number and width of such jets depend, in a non-trivial way, on the
strength of the Coriolis force and the degree of turbulence as
measured through its Reynolds number. While similar zonal jets are
present in the oceans, they tend to be less prominent than their
atmospheric counterparts due to the relatively lower levels of
turbulence~\cite{Sokolov_2007, Berloff_2011,Ocean, Ocean_Berloff, Maximenko_2005, Maximenko_2008,Cravatte_2017}.

All of this suggests not just the ubiquity of zonal turbulence in
nature, but significantly its importance in the context of geophysics---particularly in understanding large scale phenomena related to the oceans and the atmosphere. Therefore, rightly, the focus over the past few decades has
been in both understanding and modeling the barotropic quasi-geostrophic
turbulence --- the associated instabilities and formation of jets ---
and what it implies for atmospheric and oceanic patterns and
circulation~\cite{Gurcan_2015, Bouchet_2013, Bouchet_2018}.
A statistical mechanical approach serves as a powerful framework for analyzing such flows, and provides deep insight into the underlying flow behaviors~\cite{Bouchet2012}. Interestingly, fluctuations in zonal jets---an intriguing aspect of these flows---can also be characterized from the viewpoint of statistical physics. These jets manifest as bands with sharp interfaces in the vorticity configuration. Observations, both from several numerical simulations and, for example, in images of planetary flows, indicate that these interfaces---delineating regions of contrasting vorticity---also undergo significant dynamical fluctuations reminiscent of various physical phenomena. An efficient contour-tracking algorithm developed below to capture these interfaces suggests that they correspond to the velocity maxima of eastward jets---thus exhibiting the same characteristic dynamical features as the jets themselves.
 Indeed, even within the more limited context of
turbulent or turbulent-like flows, the study of such fluctuating interfaces 
have been the subject of a large body of work in recent
times~\cite{heus2008subsiding,westerweel2009momentum,chauhan2014turbulent,watanabe2015turbulent,borrell2016properties,elsinga2019turbulent,nair2021lagrangian,pal2016binary,padhan2023activity,mukherjee2025,pocheau1994scale,xin2000front,koudella2004reaction,corwin2012kardar,bentkamp2022statistical,roy2023small}.

In this paper, we focus on characterising these vorticity interfaces and show,
from detailed direct numerical simulations (DNSs) and an interface tracking
algorithm, how such interfaces are actually multifractactal with non-trivial
dynamics, non-Gaussian fluctuations while having fat-tailed distributions in
the fluctuations \textit{speeds}. We also calculate the moments of the
time-increments of the interfacial height, measured from their \textit{mean}
contour, to find that the temporal evolution is indeed rough. Furthermore, we
underline the universal aspects of such interfaces while highlighting their sensitivity to the latitude at which these observations are made.

Planetary atmospheres and oceans can be considered as \textit{shallow} fluid
layers which is a consequence of very large aspect ratio: they have a large
horizontal spread compared to their vertical depth.  Hence to unearth the
large-scale dynamics in such systems where spontaneous jet formation becomes
apparent, a useful approximation is modeling them as a homogeneous layer of
fluid on a sphere~\cite{Galperin, Scott_2007}. Thus, ignoring stratification, such thin fluids, at a given
latitude $\theta$, lend themselves to a mathematical description in terms of
the two-dimensional (2D) Navier-Stokes equation for the velocity field ${\bf
u}$ subject to a Coriolis force $2\bf{\Omega} \times {\bf u}$ with $\bf{\Omega}$
being the angular velocity of the rotating planet~\cite{Williams,
VallisMaltrud, Danilov}.  The effect of the Coriolis force at latitude $\theta$
is thus felt only through the vertical component of planetary vorticity $f =
2\Omega \sin \theta$. 

Two-dimensional fluid layers typically straddle a meridional
distance $y$ around a given latitude $\theta_0$. We account for
this spread,  within the so-called $\beta$-plane approximation~\cite{Rhines_1975}, by
assuming $y$ to be small. Thus, the net Coriolis parameter, up to a 
first-order correction, is given by $f = f_0 + \beta y$,  where $f_0 =  2\Omega \sin \theta_0$ and $\beta =  \left.\frac{\partial f}{\partial y}\right \vert_{\theta_0} = \frac{2\Omega \cos(\theta_0)}{a}$ with $a$ being the radius of the planet.

With this $\beta$ plane approximation in place, the simplest model for zonal flows would be a two-dimensional, barotropic (depth-independent),  incompressible 
Navier-Stokes equation augmented by the Coriolis parameter~\cite{Danilov, Galperin, VallisMaltrud, Williams}. 
Hence, we obtain the equation of motion for the 
vorticity $\omega =  \nabla^2 \psi$ (where $\psi$ is the stream function) field for such incompressible, 2D planetary flows:
\begin{equation}
	\label{eq:vort}
	\frac{\partial \omega}{\partial t} + J(\psi, \omega) + \beta \frac{\partial \psi}{\partial x} = \xi - \mu \omega + \nu \nabla^{2} \omega;
\end{equation}
with the Jacobian  $J(\psi, \omega) \equiv \partial_x \psi \partial_y \omega - \partial_y \psi \partial_x \omega$. 
Like for the usual two-dimensional turbulence problem~\cite{Lilly_1969, Chen_2006, Boffetta_2012, RayPRL2011}, we introduce dissipative terms through the kinematic viscosity $\nu$ 
and the coefficient of an Ekmann friction $\mu$. 
Turbulence is generated and sustained through a large scale, zero mean but delta-correlated in time, drive $\xi$ which maintains the flow in a non-equilibrium 
statistically steady state~\cite{Maltrud_Vallis_1991, Scott_2007, Scott_2012}. 

\begin{figure*}
	\centering
	\includegraphics[width=\linewidth]{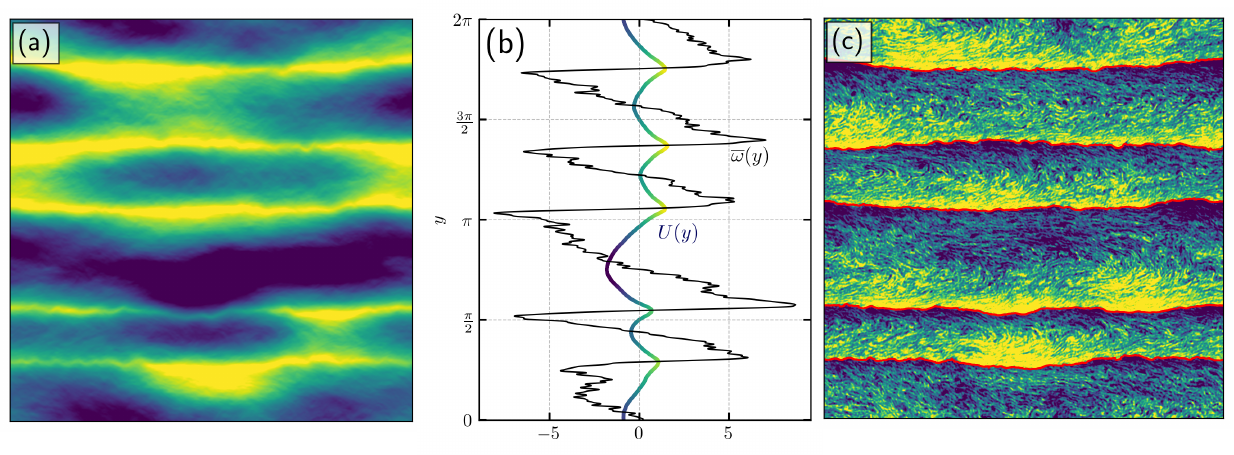}
	\caption{(a) A representative snapshot of the zonal velocity field $u(x,y)$, (b) the corresponding zonally averaged 
		mean vorticity $\overline{\omega}(y)$ and velocity $U(y)$, and (c) the vorticity field $\omega(x,y)$ 
		(like in Fig.~\ref{fig:illus}) for $\beta = 50$ ($R_\beta = 7.5$) with the interfacial contours overlaid on it. 
		The contours $\mathcal{C}$, drawn in red, separating successive shear zones 
		suggest the fluctuating nature of the interfaces whose rich dynamics is best seen in an animation of the time evolution 
		of the vorticity field~\protect\cite{YT2}.}
	\label{fig:contour}
\end{figure*}

We perform direct numerical simulations (DNSs) of Eq.~(\ref{eq:vort})
by using a standard 2/3 de-aliased pseudo-spectral algorithm on a
square periodic domain of length $L = 2 \pi$ with $N^2$ collocation
points. We use $N = 1024$ and $2048$; our results are consistent across
the two system sizes and the results presented are from the larger
simulation.  The time evolution is carried out by using a second-order,
exponential Runge-Kutta method \cite{COX2002430} with a time-stepping
$\delta t = 0.001$. We use a coefficient of friction $\mu = 10^{-3}$
and, in the DNSs for numerical stability, use a hyperviscous term
(instead of the viscous dissipation term in Eq.~\eqref{eq:vort})
$(-1)^{n+1} \nu \nabla^{2n} \omega$ with $n = 4$ and $\nu = 10^{-17}$.
The zero mean, delta-correlated in time forcing $\xi$, applied on
narrow-band of wavenumbers~\cite{Maltrud_Vallis_1991, Srini, cope_2020} around $k_f = 32$, is chosen to ensure an
energy injection rate $\epsilon = - \langle \xi \psi \rangle =
10^{-4}$. Care is also taken to ensure that no modes are forced which are
strictly  zonal $\mathbf{k}_f = (0,k_f)$ or only meridional
$\mathbf{k}_f = (k_f,0)$.

The most critical element in this study is of course $\beta$. The zonal flows,
which emerge from solutions of this two-dimensional, forced-dissipative
$\beta$-plane system, are distinct in their pattern which are characterised by
alternating bands of thin and thick jets --- the easterlies and westerlies ---
with alternating signs of velocity. In the vorticity configuration, these
manifest as banded structures with differing vorticities.  The number,
sharpness, and indeed the visual representation of such jets depend on the
latitude---with increasing $\beta$ they proliferate with sharper interfaces.
One useful measure in this context is the \textit{zonostrophy parameter}~\cite{Galperin}
\begin{equation}
	R_\beta \equiv \frac{L_{Rh}}{L_\epsilon} = \frac{1}{\sqrt{2}} U^{1/2} \beta^{1/10} \epsilon^{-1/5} 
\end{equation}
where the Rhines length scale  $L_{Rh} = \sqrt{2U/\beta}$ sets the
typical meridional widths of the zonal jets and $L_\epsilon = 2 \left (\epsilon/\beta^3 \right )^{1/5}$ 
marks the scale where transition from the homogeneous and isotropic turbulence to the one
dominated by the Coriolis parameter takes place; $U$ is the root-mean-square
velocity. 
In our DNSs, we vary 
$\beta \in \{10,20,30,40,50\}$ and cover a range of the zonostrophy parameter, 
$6.3 \leq R_\beta \leq 7.5$. Our choice of the zonostrophy parameters is consistent 
with observations in the gas giant planets~\cite{galperin2019zonal}.

As an illustrative example of the sensitive dependence of zonal turbulence on $R_\beta$, in Fig.~\ref{fig:illus} we present 
snapshots of the vorticity field for (a) $R_\beta = 6.3$ ($\beta = 10$) and (b) $R_\beta = 7.5$ ($\beta = 50$). Clearly, more distinct 
interfaces between vorticity shear zones---and hence prominent jets---form 
for the flow with the larger $R_\beta$. This is because, as is known from several studies in the past~\cite{Galperin, Sukoriansky_2007,Scott_Dritschel_2012, cope_2020}, 
with increasing $R_\beta$ a large fraction of the turbulent kinetic energy
gets organized within the mean flow which, in turn makes them stronger compared to the less coherent eddies and 
ultimately leads to steady jets. Conversely, when $R_\beta$
is small, the influence of the eddies prevail, leading to weaker, highly
fluctuating, and often indistinct zonal jets. This framework accounts for the pronounced
variability of jets in the Earth's atmosphere and oceans~\cite{ Kamenkovich_2009} (where $R_\beta$ tends to
be small) compared to the more persistent and robust jets observed in
the atmosphere of gas giant planets~\cite{Galperin_2014, Read_2024}. As a result, the zonostrophy parameter is often used as a measure of the strength, variability, and fluctuations of the emergent zonal jets~\cite{Galperin, Sukoriansky_2007}.

The spatio-temporal evolution of these banded structures, as shown in our animation of these
systems~\cite{YT1} and already suggestive in Fig.~\ref{fig:illus}, underlines a
richness in the dynamics and fluctuations of the interface between vorticity
shear zones. While the boundaries  separating regions of differing vorticity, as
seen in Fig.~\ref{fig:illus}, are visually apparent, identifying them
by means of some suitable measure requires additional effort.
Hence we develop an algorithm, as outlined below, to track the
contour $\mathcal{C}(x,t)$ --- the interface --- marking the boundary
between successive shear zones.

\begin{figure*}
	\centering
	\includegraphics[width=1.0\linewidth]{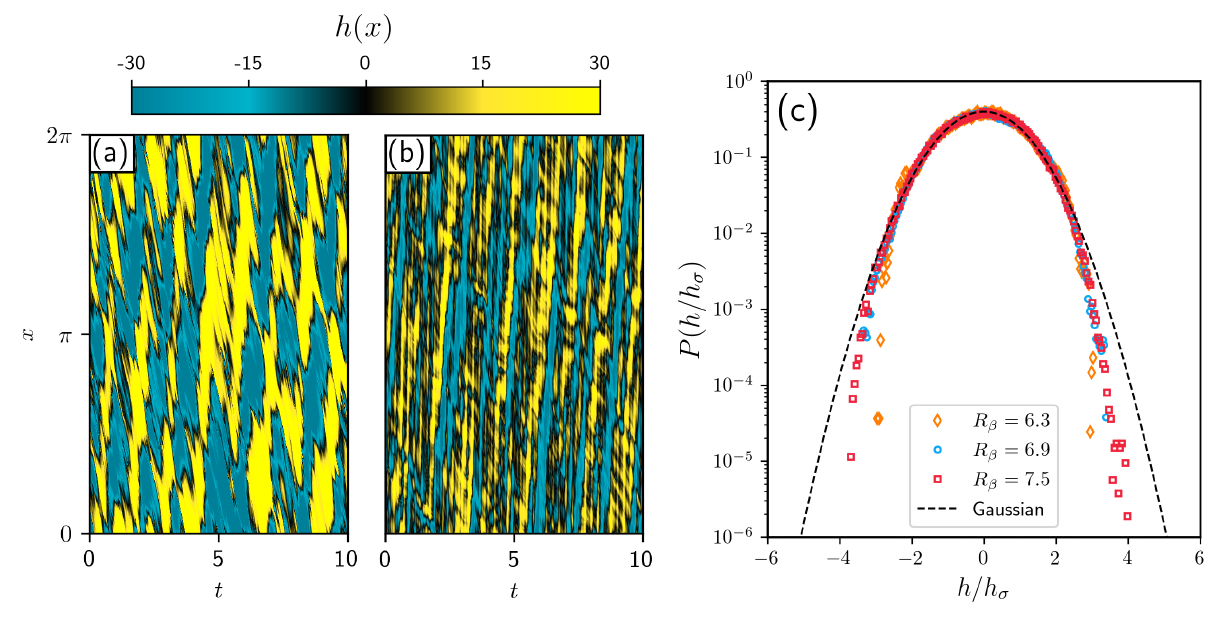}
	\caption{Representative space-time plots (kymograph) of the height field $h(x,t)$ of a particular interface for (a) $R_\beta = 6.3$ ($\beta = 10$) and 
		(b) $R_\beta = 7.5$ ($\beta = 50$). Clear diagonal banding, more prominent for smaller 
		$R_\beta$ show the conjectured lateral \textit{spilling} of the height field. (c) The 
		probability density functions (PDF) $\mathcal{P}$ of (normalised) $h/h_\sigma$ 
		for different values of $R_\beta$ $(= 6.3, 6.9, 7.5)$ (see legend), while having a Gaussian core (black dashed line),  
		show clear sub-Gaussian tails which become less pronounced with increasing $R_\beta$.}
	\label{fig:stationarystats}
\end{figure*}

The vorticity profile (Fig.~\ref{fig:illus}) consists of banded structures dominated by
either positive or negative vortices. This makes the flow vorticity
field $\omega (x,y,t)$ a natural choice for identifying the jet
boundaries since sharp transitions are expected at these locations.
However, we find that $\omega (x,y,t)$ is not a particularly useful
measure because strong fluctuations inherent in the turbulent flow
ensure that jumps are hardly discrete. Alternatively, one can examine
the zonal velocity profile, seen in Fig.~\ref{fig:contour}(a), where these
banded structures are also clearly evident. By plotting the zonally
averaged vorticity $\overline{\omega}(y,t) = \frac{1}{L}\int_0^{L}
dx \omega (x,y,t)$ and zonal mean zonal velocity $U(y,t) =
\frac{1}{L}\int_0^{L} dx \, u (x,y,t)$ as functions of $y$ at a fixed
time, we observe that near the jet boundaries the sharp jumps in
$\overline{\omega}(y,t=t_0)$ tend to coincide with the local maxima of
$U(y,t=t_0)$, as illustrated in Fig.~\ref{fig:contour}(b). This
observation motivates the use of the zonal velocity field $u(x,y,t)$ to
identify and analyze the jet interfaces.  

It appears that the vorticity interfaces correspond to the local maxima of
$u(x,y,t)$, when plotted as a function of  $y$ at a fixed value of
$x=X$ and a given time $t=t_0$. By repeating this process over the domain and across different 
snapshots,  we can systematically trace
the interfacial contour $\mathcal{C}(x,t)$ spatio-temporally. It is quite remarkable to notice from an animation~\cite{YT3}, how well the contours obtained by identifying the velocity maxima of eastward jets tracks the vorticity interfaces spatio-temporally. A snapshot of the
vorticity field with the interfacial contours overlaid as red lines
is shown in Fig.~\ref{fig:contour}(c). We refer the reader to an animation of the 
vorticity field~\cite{YT2}, overlaid with the contours $\mathcal{C}$, 
which show the intricate details in the dynamical evolution of vorticity interfaces and 
how well our algorithm tracks these fluctuating interfaces.

One of the defining characteristics of these contours is
that they exhibit spatio-temporal oscillations while maintaining their
geometric structure. The simplest way to quantify such fluctuations is
by introducing a height field $h(x,t)$, defined as the deviation of a
contour $\mathcal{C}(x,t)$ from its time-averaged profile
$\mathcal{\overline{C}}(x)$, i.e., 
\begin{equation} h(x,t) =
	\mathcal{C}(x,t) - \mathcal{\overline{C}}(x), \hskip 1em
	\text{with} \hskip 1em\mathcal{\overline{C}}(x) = \frac{1}{T}
	\int_0^T dt \, \mathcal{C}(x,t) 
\end{equation} 

What, then, is the statistics of these interfaces as captured by the
height field $h(x,t)$?  The strong east-west orientation of the zonal
velocity field allows the fluctuations to have a tangential component.
This is indeed confirmed in our measurements of the
height field and brought out succinctly in the representative kymograph plots for 
(a)$R_\beta = 6.3$ and (b) $R_\beta
= 7.5$ in Fig.~\ref{fig:stationarystats}.  The diagonal bandings, indicative of
lateral spreading of the jet fluctuations, emerge quite convincingly from such plots.  Furthermore, as $R_\beta$
increases and the jets become stronger, this lateral spreading becomes
less pronounced, as is evident from the reduced thickness of the
banding patterns when comparing panels (a) and (b) in Fig.~\ref{fig:stationarystats}.

\begin{figure*}
	\centering
	\includegraphics[width=1.0\linewidth]{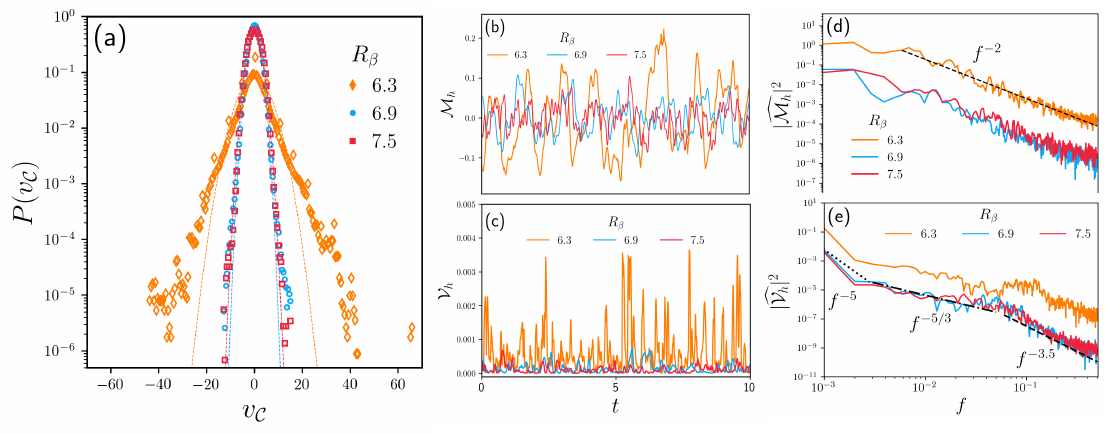}
	\caption{(a) Probability density functions (PDF) $P$ of the \textit{speed} 
		$v_{\mathcal{C}}$ of the height field for various $R_\beta$ (see legend). The dashed-curves are separate Gaussian fits to each PDF. 
		A representative time-series of the sectoral (b) mean height $\mathcal{M}_h$ and (c) 
		variance $\mathcal{V}_h$ fluctuations, 
		for a random sector of an interface, have been shown. 
		Loglog plots of the power spectrum of the (d) mean height $|\widehat{\mathcal{M}_h}|^2$ and (e) 
		variance $|\widehat{\mathcal{V}_h}|^2$, averaged over all sectors, 
		show clear power-law behaviour indicated by the black lines, which seem robust within 
		the range of $R_\beta$ considered in this study.}
	\label{fig:spacetimestats}
\end{figure*}

This lateral spread also hints at the emergence of
possible non-Gaussian tails in the probability density function
(PDF) $P(h/h_{\sigma})$, where the height field $h$ is normalised by
its standard deviation $h_\sigma$. 
By using 1000 temporally well-separated and statistically independent
snapshots of the vorticity field, we compute this PDF, as shown in
Fig.~\ref{fig:stationarystats}(c), and observe a Gaussian core
accompanied by clear sub-Gaussian tails which become less pronounced as
$R_\beta$ increases. While the numerical evidence for these sub-Gaussian 
tails is compelling, it is hard to understand from the equations of motion what could 
be the possible reasons for such a behaviour. 

Such sub-Gaussian tails are less common in measurements in fully developed turbulence. 
Nevertheless, we would still 
expect signatures of fat tails and intermittency in 
the \textit{velocity} of 
the interface --- $v_{\mathcal C}({\bf x},t) = \lim_{\delta t \to 0} \delta h/\delta t$ -- which ought to be pegged 
to the underlying zonal turbulence that result in such dynamics. 
We begin by measuring the PDF $P(v_{\mathcal{C}})$ of the speeds and find that 
(Fig.~\ref{fig:spacetimestats}(a)), contrary to the height
distribution, fluctuations in speed are indeed fat-tailed and reminiscent of fully developed turbulence. 
For smaller values of the zonostrophy parameter, the eddies and the mean flow compete as discussed earlier resulting in strongly 
intermittent interfacial velocities. However, 
as the zonostrophy parameter becomes large, the mean flow dominates and hence it seems likely that intermittency ought to weaken. 
In Fig.~\ref{fig:spacetimestats}(a), we find this heuristic argument to be reasonably accurate with a limiting Gaussian behaviour 
for the larger $R_\beta$.

We now look at the spatiotemporal properties of $h(x,t)$ and quantify
the timescales of fluctuations. We find it convenient to address this issue 
by understanding the sectorial mean and variance of the interface.
These sectors are obtained by segmenting
the interfacial contour $\mathcal{C}(x,t)$ into $N_s$ ($=32$ for our
analysis) bins. Thus the sectoral mean interfacial height
$\mathcal{M}_{h,s}(t)$ and variance $\mathcal{V}_{h,s}(t)$ are defined as 

\begin{equation}
	\begin{gathered}
		\mathcal{M}_{h,s}(t) = \frac{1}{N_p}\sum_{x \in s} h(x,t), \\
		\mathcal{V}_{h,s}(t) = \frac{1}{N_p} \sum_{x \in s} (h(x,t) - \mathcal{M}_{h,s}(t))^2
	\end{gathered}
\end{equation}
where $N_p$ ($=N/N_s$) is the number of points in each sector. The temporal
statistics is performed on 1000 evenly-spaced time snapshots obtained in the
statistically steady state. 

The time series of $\mathcal{M}_{h,s}$ and $\mathcal{V}_{h,s}$, shown in
Fig.~\ref{fig:spacetimestats}(b) and (c) for a representative, random segment, exhibits large fluctuations which seem to depend, 
in magnitude, on the value of $R_\beta$. The time series of $\mathcal{M}_{h,s}$ is strongly reminiscent of the saw-tooth solutions of  
the Burgers equation and  
suggests that 
the spectrum of the mean height $|\widehat{\mathcal{M}_h}|^2$ decaying as $f^{-2}$ at high frequencies $f$. To confirm a possible 
Burgers-like dynamics, we measure this spectrum averaged over all $N_s$ sectors and indeed find a clear $f^{-2}$, as shown in 
Fig~\ref{fig:spacetimestats}(d). 

\begin{figure*}
	\centering
	\includegraphics[width=1.0\linewidth]{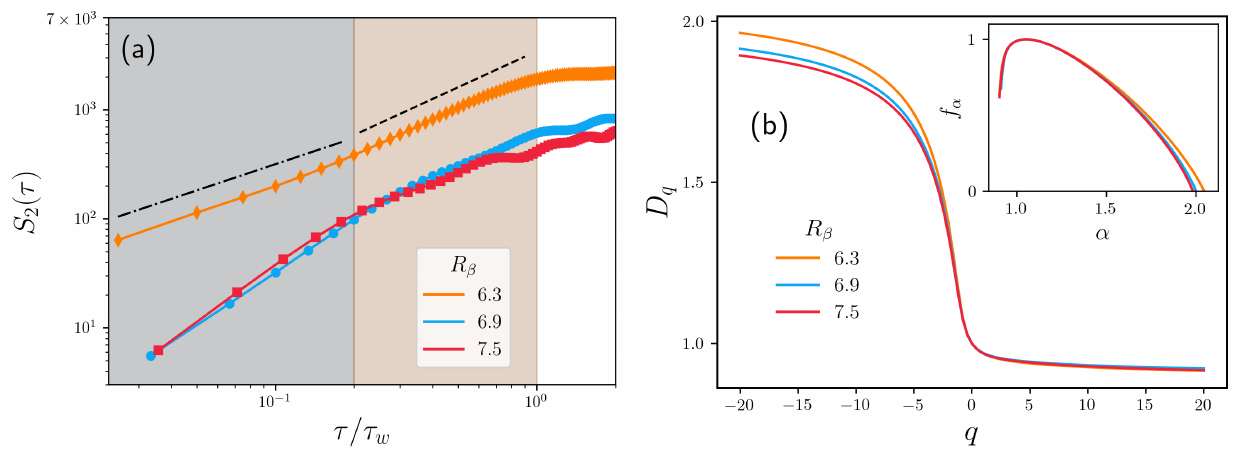}
	\caption{(a) A loglog plot of the second-order structure function $S_2$ versus 
	the normalised time $\tau/\tau_w$ showing a power-law regime with an exponent $\xi_2$, 
	for $\frac{\tau}{\tau_w} \lesssim 1$, as indicated by the dashed line. At short times, 
	the function shows a non-ballistic behaviour with a $R_\beta$-dependent 
	exponent $\zeta_2 \neq 2$ as shown by the dashed-dotted line.
	(b) Plots of the generalised dimensions $D_q$ vs $q$ and (inset) the singularity 
		spectrum $f(\alpha)$ vs $\alpha$ for different values of $R_\beta$.}
	\label{fig:MF}
\end{figure*}

The time series of $\mathcal{V}_{h,s}$, a second-order quantity modelling the interfacial energy, on the other hand, is 
more \textit{intermittent}. Indeed, it is reasonable to expect that the corresponding frequency spectrum ought 
to reflect the energy spectrum, especially for large $R_\beta$, 
of the underlying turbulent zonal flow which triggers such interfaces. In Fig~\ref{fig:spacetimestats}(e), we show 
plots of the spectra of the height variance (averaged over all sectors)  $|\widehat{\mathcal{V}_h}|^2$ and find compelling evidence 
of the distinct power law regimes --- $f^{-5}$, $f^{-5/3}$, and $f^{-7/2}$ --- characteristic of $\beta$-plane 
turbulence. Furthermore, this contrasts effectively with the variance of height fluctuations seen in the arrested 
growth of interfaces marking the separation of binary active fluids reported recently~\cite{mukherjee2025} where 
the interface provides an effective two-dimensional turbulent annular shell separating the more active 
fluid from the less active one.

The time-series of the height fluctuations --- $\mathcal{M}_h$ and $\mathcal{V}_h$ --- are suggestive. 
While they appear self-similar --- confirmed by the power-laws in the corresponding spectrum --- the 
sharp, dense, \textit{shock}-like features suggest a possible lacking of differentiability. In particular, 
while the intermediate, \textit{inertial} range statistics may be H\"older continuous with $h < 1$, it is 
not obvious that the height field $h(t)$ (integrated over space) is smooth and differentiable. In order to 
test these ideas, we calculate the second-order \textit{structure function} of the height field, with 
time increments $\tau$ conveniently normalised by the Rossby time-scale $\tau_w = \nicefrac{1}{L_\epsilon\beta}$, 
\begin{equation}
	S_2(\tau) = \langle \left [h(t+\tau) - h(t)\right ]^2 \rangle \sim 
	\begin{cases} 
		\left( \frac{\tau}{\tau_w} \right)^{\zeta_2}, \quad \quad \text{for $\frac{\tau}{\tau_w}  \ll 1$} \\
		\left( \frac{\tau}{\tau_w} \right)^{\xi_2}, \quad \quad \text{for $\frac{\tau}{\tau_w} \lesssim 1$}
	\end{cases} 
\end{equation}
with the scaling exponent $\zeta_2 = 2$ for differentiable, smooth functions and $S_2$ saturating 
for $\tau/\tau_w > 1$. 

In Fig.~\ref{fig:MF}(a), we plot this second-order temporal structure function, on a log-log scale, 
for three different values of $R_\beta$. Clearly there does seem to be an \textit{inertial} range, shaded in 
beige with the dashed black line as a guide to the eye, 
with $\xi_2 \approx 1.1$. Furthermore, the dependence 
of this exponent $\xi_2$ on $R_\beta$ seems to be mild and the variation insignificant within error bars. At times of the order of the Rossby time, the second-order structure function saturates.

But what stands out in Fig.~\ref{fig:MF}(a) is the scaling behaviour when $\tau/\tau_w \ll 1$. Remarkably, and as anticipated 
earlier, $\zeta_2 \neq 2$ underlining the roughness of the temporal fluctuations of the height field. Furthermore, 
$\zeta_2 \to 2$ as $R_\beta$ increases. This numerical observation must serve as a crucial ingredient in future mathematical modelling through possible stochastic differential equations of the contour $\mathcal{C}$. 

Before we conclude, we make a final characterisation of the nature of fluctuations to complete our study.
A wide range of numerical and experimental studies has pointed out that
several aspects of fully developed turbulence, such as the turbulent
energy dissipation field, are fractals and are best characterized by a broad range
of self-similar scales~\cite{Sreenivasan_Meneveau_1986, PhysRevLett.59.1424, Meneveau_Sreenivasan_1991}. As another quantification of the interface
fluctuations, we carry out a multifractal analysis of the height field~\cite{mukherjee2025}.
In doing so, we define the function $\mathcal{H}(x,y,t) = |h(x,y,t)| \,
+ \, 0.001 $ (here, $0.001$ is added to the absolute value of the
height field to make $\mathcal{H}(x,y,t)$ positive definite). Then we
follow the same procedure in constructing the partition function of
$\mathcal{H}(x,y,t)$ as discussed in many earlier studies~\cite{MENEVEAU198749, Frisch_1995, MukherjeePRL2024}.  We
divide the interfacial contour into $N_l$ segments of size $l$ (where
$N_l = L/l$, with $L$ being the domain length in the simulation and $l
\in \{2^0, 2^1, 2^2,...,2^{11} \}$ points), and find the coarse-grained
height at this scale for each partition, like for the $i$--th partition
it is given by $\mathcal{H}_{l,i} = \sum_{x = li}^{l(i+1)}
\mathcal{H}(x,y)$ where $i \in [0, \, N_l - 1]$. Then we add powers of
different order $q$ (real) of $\mathcal{H}_{l,i}$ over all segments to
get the partition function $Z_q(l)$ ($ \equiv \sum_{i=0}^{N_l -1}
\mathcal{H}_{l,i}^q$), and the process is repeated for different values
of $l$. If the height field is multifractal, we expect $Z_q(l)$ to
scale with the size of the partitions $l$ according to some power law.
In this context, the so-called generalized dimensions $D_q$ are
defined, such that $Z_q(l) \sim l^{(q-1)D_q}$. The plots of $\ln
Z_q^{\frac{1}{q-1}}$ vs $\ln l$ will have a linear region in the
inertial range, whose slope gives the $D_q$. For different values of
$q$, we get a distribution of $D_q$ vs $q$. Numerically we obtain the
$D_q$ vs $q$ curve by ensemble averaging $D_q$ over 1000 snapshots of
the interfacial height. From this we arrive at the singularity spectrum
$f_\alpha -\alpha$ via the Legendre transform of $(q-1)D_q$, as $\alpha
= \frac{d}{dq}(q-1)D_q$ and $f_\alpha = \alpha q - (q-1)D_q$. 

The distribution of $D_q$ vs $q$ for different values of the zonostrophy parameter are shown 
in Fig.~\ref{fig:MF}(a);  we see a more pronounced behavior towards the negative values
of $q$ and has a distinctive feature for different $R_\beta$s. The
positive $q$ part of the distribution is almost flat, and doesn't vary
as $R_\beta$ changes. This indicates that lower magnitudes of the height
profile exhibit more of the multiscale fluctuations, while larger
deviations are smoother. Also, as $R_\beta$ increases, the fluctuations
get smoother resulting in the downward trend of the distribution curve.
The associated $f_\alpha - \alpha$ spectrum of singularity strengths
$\alpha$, shown in Fig.~\ref{fig:MF}(b), exhibits a positively skewed profile with
a broad range of $\alpha$ values, indicating that the fluctuations are
characterized by a spectrum of scaling exponents. Interestingly, as
$R_\beta$ increases, the degree of multifractality diminishes. This is
also an outcome of the fact that interface fluctuations are reduced
with increasing $R_\beta$. As a result, the interfacial profile gets
smoother, resulting in an overall reduction in the range of $\alpha$.

In this work, we make a systematic characterization of the vorticity interfaces
which emerge spontaneously in two-dimensional, barotropic, $\beta$-plane
turbulence. In particular, we find that the dynamics of such interfaces are complex and exhibit multifractal statistics. Given the importance of zonal
flows in particle-laden atmospheric and marine systems, the
statistics of such interfaces, which induce a distinctive lateral shear, must
play an important role in the stretching and subsequent transport of elastic,
filamentary pollutants --- such as microplastics --- which have been shown to
have a preferential (Lagrangian) sampling of the
flow~\cite{Picardo2018,Picardo2020}. Furthermore, such interfaces may play an
important role in determining coalescence, orientation and aggregation of
sub-Kolmogorov particles~\cite{Bec2013,Bec2016,Anand2020}, essentially because
we know how fluid structures determine the fate of collision and coalescence
mechanisms in fully developed turbulence~\cite{Picardo2019}. Finally, how such jets influence Lagrangian intermittency 
and chaos~\cite{Ray2018} is left for future work.


We are very grateful to S. Mukherjee and V. Vasan for several useful discussions and suggestions. 
S.S.R. acknowledges SERB-DST (India) projects 
STR/2021/000023,  CRG/2021/002766, and the CEFIPRA Project No 6704-1 for  support. 
This research was supported in part by the International Centre for Theoretical Sciences (ICTS) for the programs - 
Field Theory and Turbulence (code:ICTS/ftt2023/12), 
Indo-French workshop on Classical and quantum dynamics in out of equilibrium systems  (code: ICTS/ifwcqm2024/12) and 
10th Indian Statistical Physics Community Meeting (code: ICTS/10thISPCM2025/04). 
The simulations were performed on the ICTS clusters Mario, Tetris, and Contra.
The authors 
acknowledge the support of the DAE, Government of India,
under projects nos. 12-R\&D-TFR-5.10-1100 and RTI4001.

\bibliography{apssamp}

\end{document}